\newcommand{\rem}[1]{}
\documentclass[12pt]{iopart}
\usepackage{graphicx}
\usepackage{iopams} 

\newtheorem{thrm}{Theorem}[section]
\newtheorem{mainthrm}{Main Theorem}[section]
\newtheorem{lem}[thrm]{Lemma}
\newtheorem{prop}[thrm]{Proposition}
\newtheorem{cor}[thrm]{Corollary}
\newtheorem{remark}[thrm]{Remark}

\renewcommand\theenumi{\roman{enumi}}

\begin{document}
\title[Identities and exponential bounds]
{Identities and exponential bounds\\ for tranfer matrices}

\author{Luca~Guido~Molinari}
\address{Physics Department\\
Universit\'a degli Studi di Milano and I.N.F.N. sez. Milano\\
Via Celoria 16, 20133 Milano, Italy}
\ead{luca.molinari@mi.infn.it}

\begin{abstract}
This paper is about analytic properties of single 
transfer matrices originating from general block-tridiagonal or 
banded matrices.
Such matrices occur in various applications in physics and numerical analysis.
The eigenvalues of the transfer matrix describe localization of eigenstates 
and are linked to the spectrum of the block tridiagonal matrix 
by a determinantal identity. If the block tridiagonal matrix is invertible, 
it is shown that half of the singular values of the transfer matrix have a 
lower bound exponentially large in the length of the 
chain, and the other half have an upper bound that is exponentially small. 
This is a consequence of a theorem by Demko, Moss and Smith on the decay of 
matrix elements of inverse of banded matrices. 
\end{abstract}
\submitto{\JPA}
\maketitle

\section{Introduction}
Consider the difference equation 
\begin{equation}
C_k u_{k-1}+A_ku_k +B_ku_{k+1}= E u_k \, , \qquad  k=1\ldots n, \label{chain}
\end{equation}
where $A_k,\, B_k,\, C_k\in \mathbb C^{m\times m}$ are complex non 
singular square matrices, $E$ is a complex parameter, 
and $u_k\in \mathbb{C}^m$ are unknown vectors. It is the prototype of
several equations that occur in physics.\\
At each $k$ the equation provides $u_{k+1}$ in terms of $u_k$ 
and $u_{k-1}$; the recursion is made single-term by doubling the vector
and introducing the 1-step transfer matrix $t_k(E)$, of size $2m$:
\begin{eqnarray}
\left [ \begin{array}{c} u_{k+1}\\ u_k\end{array}\right ] = 
\left [\begin{array}{cc} B_k^{-1}(E-A_k) & -B_k^{-1}C_k\\ 
\mathbb I_m & 0 \end{array}\right ]
\left [ \begin{array}{c} u_k \\ u_{k-1}\end{array}\right ].
\label{transferk}
\end{eqnarray}
Iteration builds up the $n$-step
transfer matrix $T(E)=t_n(E)\cdots t_1(E)$ that connects vectors $n$ 
steps apart:
\begin{eqnarray}
 T(E) \left [ \begin{array}{c} u_1 \\ u_0\end{array}\right ]=
\left [ \begin{array}{c} u_{n+1}\\ u_n\end{array}\right ].
\label{transfer}
\end{eqnarray}
One is often interested in the singular values
$\sigma_1\ge \ldots \ge\sigma_{2m}$ of $T(E)$ (the eigenvalues of the positive
matrix $(T^\dagger T)^{1/2}$), which describe the growth or 
decay of $\|u_n\|$. 
The product of the $p$ largest ones ($p=1,\ldots ,2m$) can be 
obtained by the formula
$\sigma_1\cdots \sigma_p = \| \Lambda^p T(E)\|$, where  
$(\Lambda^pT)(v_1\wedge\ldots \wedge v_p) =: Tv_1\wedge \ldots \wedge Tv_p$
extends the action of $T$ to antisymmetric $p-$forms and $\|O\|$ is
the sup norm of operators \cite{Winitzki,Bhatia}. For real transfer 
matrices the product has the simple geometric interpretation 
\begin{eqnarray}
\sigma_1\cdots \sigma_p = 
\sup_{v_1\ldots v_p } 
\frac{{\rm Volume\; P} \{Tv_1, \ldots , Tv_p\} }
{{\rm Volume \;P} \{v_1, \ldots , v_p\} }
\end{eqnarray}
where ${\rm P}\{ v_1, \ldots ,v_p \}$ is the parallelogram with sides 
$v_i\in \mathbb R^{2m}$.\\
When the transfer matrix is the product of random matrices, 
Oseledets' Multiplicative Ergodic
Theorem ensures that the singular values grow or decay 
exponentially in $n$ with rates (Lyapunov exponents) 
$\lambda_k =\lim_{n\to\infty} \frac{1}{n}\ln \sigma_k$
that are independent of the realization \cite{Bougerol, Crisanti}. 
Then: 
$$ \lambda_1+\ldots +\lambda_p = \lim_{n\to\infty} 
\frac{1}{n}\ln \|\Lambda^p T\|. $$
The formula can be implemented numerically for the evaluation of Lyapunov
spectra \cite{Slevin}. In the simplectic case ($\lambda_{m+k}=-\lambda_k$)
the average of the positive Lyapunov exponents is expressible in terms of
the average distribution of eigenvalues of the Hermitian random matrices
associated to \eref{chain}:
\begin{equation}
 \frac{\lambda_1+\ldots +\lambda_m}{m} 
= \int dE' \rho (E') \ln |E-E'| + {\rm const.}
\label{KuSoTh}
\end{equation}
The formula was obtained by Herbert, Jones and Thouless for $m=1$, 
and by Kunz, Souillard, Lacroix \cite{Lacroix} for $m>1$.
It is desirable to obtain similar equations for the evaluation of single
or other combinations of the exponents.\\

In this paper the properties of a single 
transfer matrix are investigated. It will be proven that, for large $n$, 
half of its singular values have a lower bound that grows exponentially in 
$n$, and the other half have
an upper bound that decays exponentially in $n$. Moreover, the spectrum of 
eigenvalues will be linked, via duality, to the spectrum of the difference 
equation \eref{chain} with proper boundary conditions.\\
The idea of duality is simple. 
For a chain of length $n$, 
if Bloch boundary conditions (b.c.) $u_{n+1}=e^{\rmi\varphi}u_1$ and 
$u_0=e^{-\rmi\varphi}u_n$ are chosen (they correspond to an 
infinite periodic chain), an eigenvalue equation is obtained:
\begin{eqnarray}
 T(E) \left [ \begin{array}{c} u_1 \\ u_0\end{array}\right ]=
e^{\rmi\varphi}\left [ \begin{array}{c} u_1\\ u_0\end{array}\right ].
\label{bloch}
\end{eqnarray}
The condition $\det [T(E)-e^{\rmi\varphi}\mathbb I_{2m}]=0$ 
gives the $nm$ eigenvalues $E_a (\varphi)$ of the difference equation
\eref{chain}. 
Then, for each eigenvalue, the whole eigenvector of the chain
$(u_1\ldots u_n)$ is constructed by applying the 
1-step transfer matrices to the initial vector $(u_1,u_0)$.\\  
The opposite approach is also useful. The eigenvalue equation for $T(E)$ 
\begin{eqnarray}
T(E) \left [ \begin{array}{c} u_1\\ u_0\end{array}\right ] = z
\left [ \begin{array}{c} u_1 \\ u_0\end{array}\right ]
\label{transferz}
\end{eqnarray}
is solved whenever $(u_1,\ldots ,u_n)^t$ is an eigenvector with eigenvalue 
$E$ of the matrix
\begin{equation}\label{Hz}
H (z)=\left[\begin{array}{cccc}
A_1 & B_1 & {} &  \frac{1}{\displaystyle z}C_1 \\
C_2  & \ddots & \ddots & {}\\
{} & \ddots & \ddots & B_{n-1}\\
z B_n & {} & C_n & A_n 
\end{array}\right]
\end{equation}
which encodes the b.c. $u_{n+1}=zu_1$ and $u_0=u_n/z $ that are
implied by the eigenvalue equation for the transfer matrix. The statement
\begin{prop}\label{statduality}
$(u_1,\ldots ,u_n)^t$ is a right eigenvector with eigenvalue 
$E$ of the matrix $H(z)$ if and only if $(u_1, u_n/z)^t$ is a right 
eigenvector of $T(E)$ with eigenvalue $z$,
\end{prop}
\noindent
translates into a determinantal identity (the duality relation,
 \cite{Molinari97}) that relates 
the eigenvalues of the tranfer matrix $T(E)$ to those the associated 
``Hamiltonian'' matrix $H(z)$, that describes the difference equation of length 
$n$ with generalized Bloch boundary conditions.\\

It is occasionally useful to replace the parameter $z$ with $z^n$. The matrix 
$H(z^n)$ is similar to the balanced matrix 
\begin{equation}\label{HBz}
H^B(z)=\left[\begin{array}{cccc}
A_1 & zB_1 & {} &  \frac{1}{\displaystyle z} C_1 \\
\frac{1}{\displaystyle z} C_2  & \ddots & \ddots & {}\\
{} & \ddots & \ddots & zB_{n-1}\\
z B_n & {} & \frac{1}{\displaystyle z} C_n & A_n 
\end{array}\right]
\end{equation}
by the similarity relation $ H (z^n)\,=\,D(z) H^B(z) D(z)^{-1} $, where
$D(z)$ is the block diagonal matrix $(z \mathbb I_m,\ldots, z^n\mathbb I_m)$.
As a consequence $H(z^n)$, $H^B(z)$ and also $H^B(ze^{\rmi k 2\pi/n})$, 
$k=1\ldots n-1$, have the same eigenvalues.\\ 
While the matrix $H(z^n)$ remarks the value of $z^n$ as a boundary condition
parameter, the matrix $H^B(z)$ remarks the invariance under cyclic permutations
of blocks (the ring geometry) of the difference equation (and is 
numerically more tractable).\\ 

\begin{figure}\label{figl}
\begin{center}
\includegraphics[angle=0,width=5cm]{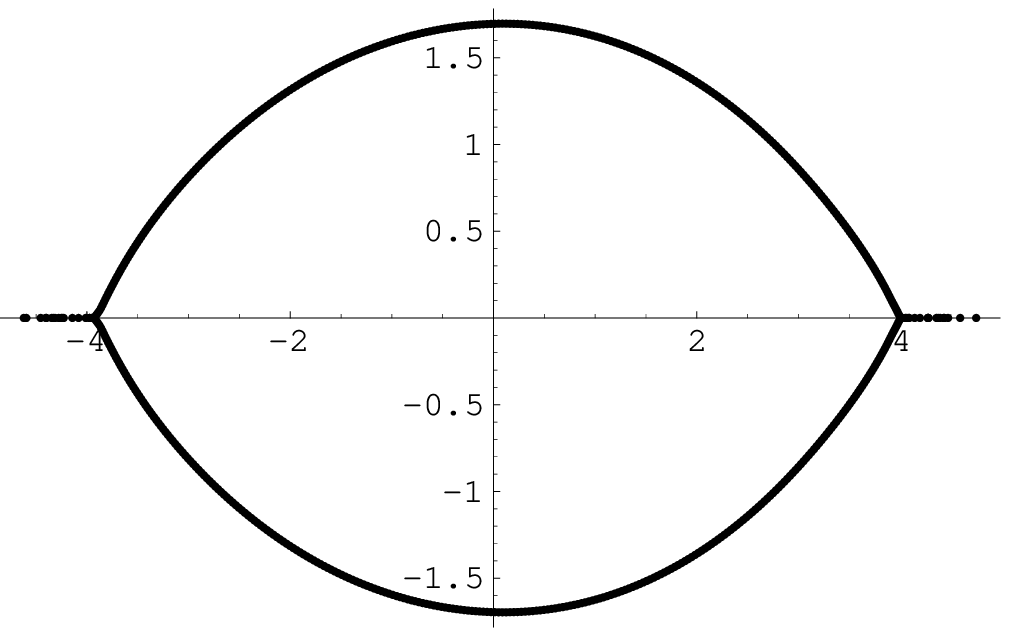}
\includegraphics[angle=0,width=5cm]{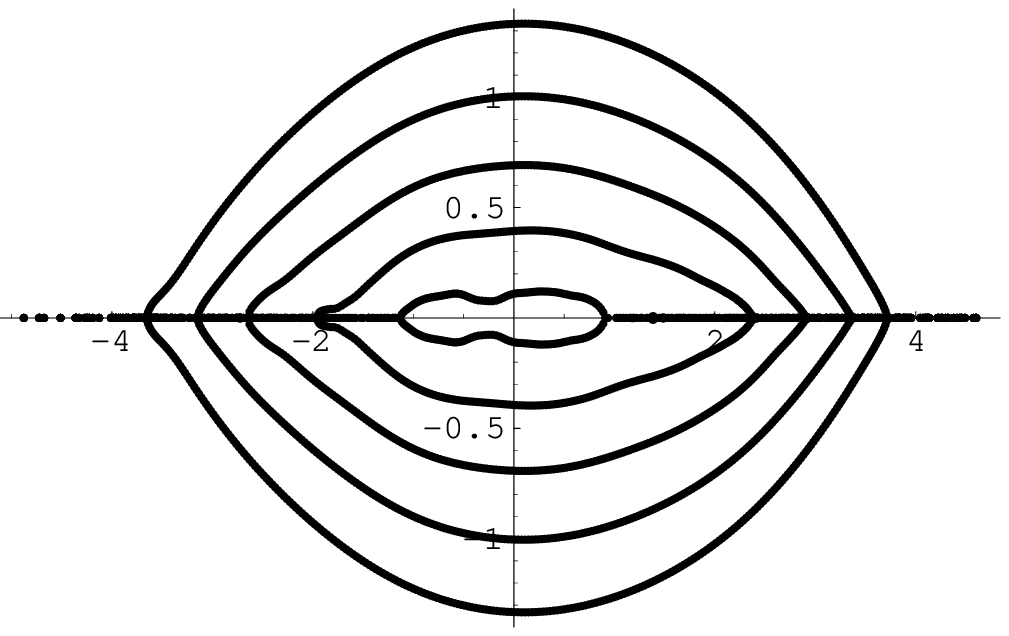}
\caption{\label{Fig1} Left: the complex eigenvalues of a Hatano Nelson
tridiagonal matrix ($m=1$, $n=600$, $\xi =1$) with random diagonal 
elements uniformly chosen in $[-3.5,+3.5]$. They lie on the line
$\xi =\lambda (E)$. The real eigenvalues correspond to states with 
localization length less than $1/\xi$.
Right: the same system, with $\xi $ increasing from $0$ to $1$ in five steps
to show the expanding spectral curve.}
\end{center}
\end{figure}

\begin{figure}\label{figlb}
\begin{center}
\includegraphics[angle=0,width=5cm]{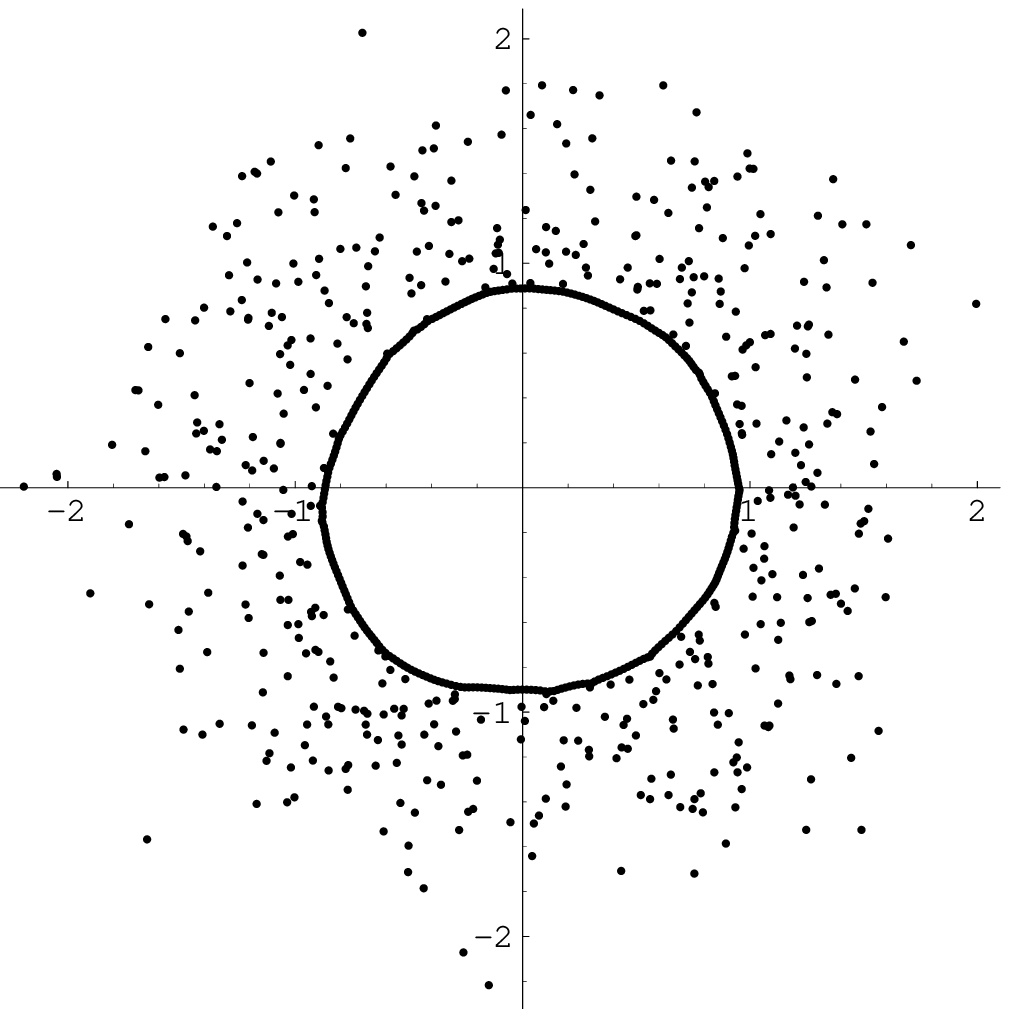}
\includegraphics[angle=0,width=5cm]{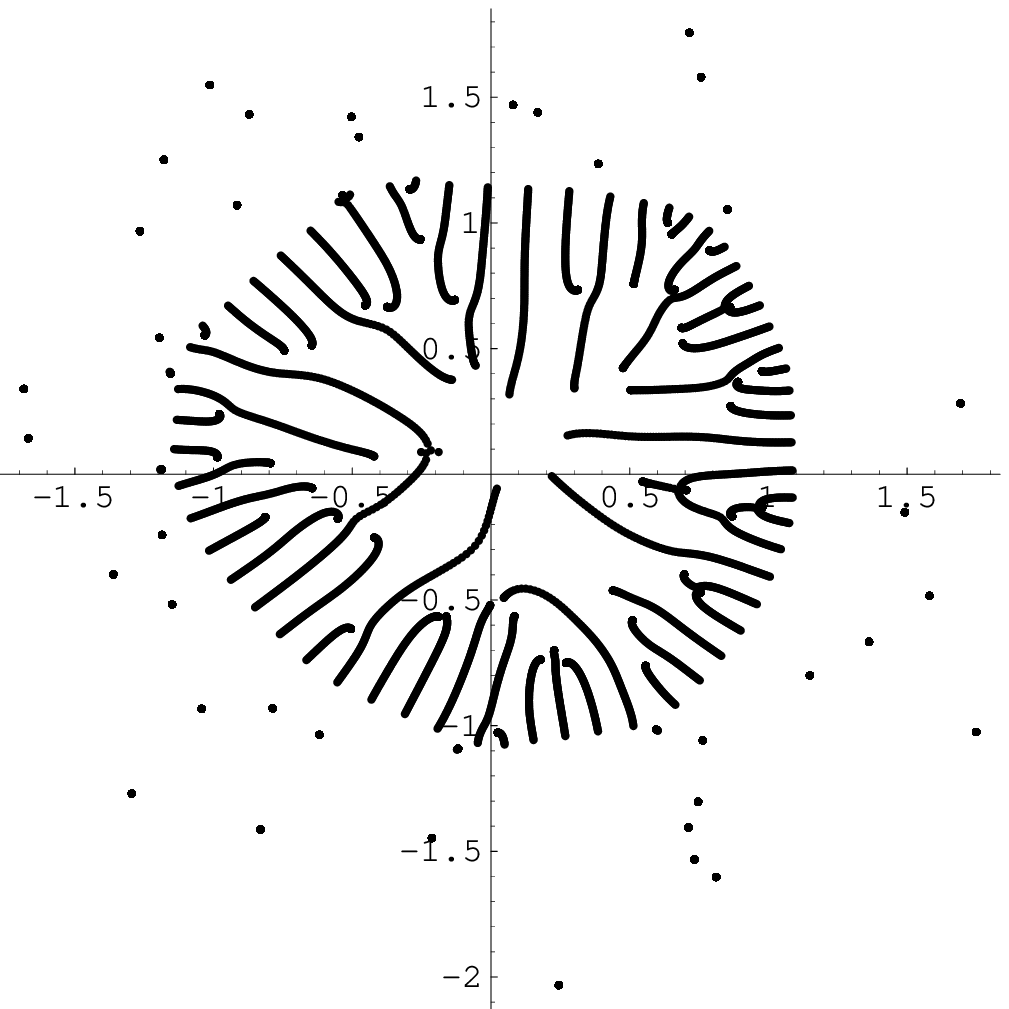}
\caption{\label{Fig1b} 
Left: the eigenvalues of a tridiagonal matrix
($m=1$, $n=800$, $\xi=0.5$) with elements $a_k,b_k,c_k$ chosen uniformly 
in $[-1,1]$; the ``front circle'' contains the eigenvalues that filled
the circle at lower values of $\xi $.
Right: the motion of eigenvalues ($n=100 $)
is traced for $\xi $ changing from $0.3$ to $0.6$. The outer eigenvalues are
numerically unaffected before being reached by the ``front circle''.}
\end{center}
\end{figure}

Tridiagonal matrices of type \eref{HBz}, with $z=e^\xi$ real, were 
introduced by
Hatano and Nelson \cite{Hatano} to model vortex pinning in superconductors:
\begin{equation*}
 e^\xi u_{k+1}+ a_k u_k + e^{-\xi} u_{k-1} = E u_k , 
\end{equation*}
where $a_k$ are independent random entries. The model attracted a 
great interest as it gave another view of the relationship between 
localization and spectral response to b.c. variations.
For zero or small $\xi$ the eigenvalues are real and all eigenvectors are
exponentially localized with localization lengths $1/\lambda (E)$. The 
Lyapunov exponent can be evaluated by Thouless' formula \eref{KuSoTh}, 
$\lambda (E)= \int dE \rho (E) \ln |E-E'|$, with the average 
spectral density of the Hermitian chain (the analytic evaluation is
possible in Lloyd's model, with Cauchy disorder \cite{Lloyd}). 
By increasing $\xi $ 
beyond a critical value the eigenvalues start to gain imaginary parts and
distribute along a single expanding curve \cite{Goldsheid00} of equation 
$\xi = \lambda (E)$ (see figure~\ref{Fig1}). The transition has been 
studied also in 2D, where the critical value of $\xi $ for the onset of
migration in the complex plane gives the inverse localization in the
center of the band \cite{Kuwae}.\\
If the parameter $\xi $ is turned on in tridiagonal random matrices that
are not Hermitian at the beginning, 
\begin{equation*}
 b_ke^\xi u_{k+1}+ a_k u_k + e^{-\xi} c_k u_{k-1} = E u_k , 
\end{equation*}
the phenomenon shows up differently \cite{Molinari09b}: beyond a critical 
value of $\xi $, an area occupied 
by the complex eigenvalues starts to be depleted, the eigenvalues being 
swept away and accumulated on an expanding 
``front line'' of equation $\xi = \lambda (E)$. No 
eigenvalues are left in the interior (corresponding to delocalization of 
states) (see figure~\ref{Fig1b}).

{\quad}\\
Though the theory presented in this paper is very general, these two 
models were the starting motivation:\\
1) {\bf band random matrices} have block tridiagonal structure 
with  lower and upper triangular $B$ and $C$ matrices. Matrix 
elements are independent and identically distributed (i.i.d.) random 
variables. It is customary to name $m$ as $b$ (bandwidth is $2b+1$).
If the probability distribution has zero mean and finite variance, and
if $n\gg b\gg 1$, the spectral density of Hermitian banded matrices 
is Wigner's semicircle law, with exponentially localized eigenvectors. The 
localization length and its finite size scaling 
were studied numerically by Casati et al.\cite{Casati}, with 
insight provided by the kicked rotor model of quantum chaos. 
Several properties were obtained analytically by supersymmetric techniques in a 
series of papers by Fyodorov and Mirlin \cite{Fyodorov}.\\
2) {\bf Anderson model} describes the propagation of a particle in a 
lattice with random site potential. After choosing a (long) 
direction of length $n$, the diagonal blocks $A_k=T+D_k$ describe the 
sections of the lattice with $m$ sites each ($T$ is the Laplacian matrix 
for the transverse slice and $D_k$ is a random diagonal matrix with i.i.d. 
elements). The hopping among neighboring slices is fixed by 
$B_k=C_k=\mathbb I_m$. The random site-potential is usually chosen
uniformly distributed in $[-w/2,w/2]$ ($w$ is the disorder parameter)
(the literature is vast, see \cite{Stolz} for a mathematical introduction).

In both models the transfer matrix is a product of random matrices and,
for $n\to\infty$, it provides a non random Lyapunov spectrum 
\cite{Kottos,Baldes,Zhang}.
The inverse of the smallest Lyapunov exponent is the localization length.\\ 
Localization affects the response of energy values to variations of b.c.
 \cite{Casati2,Zyczkowski}. 
This dual way of viewing localization: through decay of eigenvectors 
(transfer matrix) or response of 
energy levels to b.c. variations (Hamiltonian matrix), is hidden in
the duality identity among the eigenvalues of $T(E)$ and of $H(z)$.\\

The first two sections provide algebraic properties that relate a 
generic transfer matrix to its Hamiltonian matrix. Some of them appeared in 
previous papers, but receive here a consistent presentation. In particular, 
they are the spectral duality and the expression of $T(E)$ in terms of the 
resolvent of the Hamiltonian matrix with open b.c.\\ 
Next, a theorem by Demko, Moss and Smith \cite{Demko} on the decay
of matrix elements of the inverse of a banded matrix is presented. It
is used here to prove that a $2m\times 2m$ transfer matrix has $m$ 
singular values growing exponentially with the length of the chain, 
and $m$ singular values decaying exponentially. This new result reflects
on a single matrix a property of random matrix products.\\
The rest of the paper deals with identities; duality and Jensen's 
identity give an expression for the exponents $\xi_a=\frac{1}{n} \ln |z_a| $, 
where $z_a$ are the eigenvalues of the transfer matrix, in terms of the 
eigenvalues of the associated matrix $H(z)$. Hadamard's inequality 
for determinants of positive matrices supports the idea that the 
eigenvalues $z_a$ have a leading exponential growth in $n$.
The discussion of the relevant case of Hermitian difference equation ends 
the paper.

\section{Transfer matrix and duality}
Some general facts about transfer matrices are presented. By 
construction $T(E)$ is a polynomial in $E$ of degree $n$, 
$T(E)=E^n T_n+\ldots +ET_1+T_0$, with matrix coefficients. 
However, its determinant is independent of $E$:
\begin{equation}
   \det T(E) = \prod_{k=1}^n \det t_k(E) =
\frac{\det [C_1\cdots C_n]}{\det [B_1\cdots B_n]}
\end{equation}
This implies that $T(E)^{-1}$ is again a matrix 
polynomial in $E$ \cite{Gohberg}. Actually  $T(E)^{-1}$  
is similar to the tranfer matrix of the inverted chain. Let's introduce the 
two matrices of inversion, of size $2m\times 2m $ and $nm\times nm$:
\begin{equation*}
\sigma_x =:
\left[\begin{array}{cc} 0 & \mathbb I_m \\ \mathbb I_m & 0\end{array}\right],
\qquad
J = \left[\begin{array}{ccc} {} & {} & \mathbb I_m \\ {} & \ldots & {}\\ 
\mathbb I_m & {} & {} \end{array}\right],
\end{equation*}
\begin{prop}\label{prop1}
Let $T(E)$ be a transfer matrix and $H(z)$ the associated matrix, and let
$T(E)^J$ be the transfer matrix associated to $H^J(z)=JH(z)J$ 
(the inverted chain); then:  $T(E)^{-1}=\sigma_x T(E)^J\sigma_x$.\\ 
Proof: {\rm $T(E)^{-1}=[t_n(E)\cdots t_1(E)]^{-1}=
t_1(E)^{-1}\cdots t_n(E)^{-1}$. The combination
\begin{equation*}
\sigma_x\, t_k^{-1} \sigma_x=
\left[\begin{array}{cc} C_k^{-1}(E-A_k) & -C_k^{-1}B_k \\ \mathbb I_m & 0
\end{array}\right]
\end{equation*}
gives the structure of a 1-step transfer matrix. Multiplication yields
the result.} \hfill{$\square $}
\end{prop}

\begin{prop}\label{powerexpT}
In the expansion of the characteristic polynomial of the transfer matrix,
$$
\det \left[ z\mathbb I_{2m} - T(E)\right ]
 =\, z^{2m}+ \ldots +a_k(E)z^{2m-k}+\ldots 
 +a_{2m-k}(E)z^k +\ldots +a_{2m},$$
the coefficients $a_k(E)$ and $a_{2m-k}(E)$ are (in general different)
polynomials in $E$ of degree $kn$ ($k=0,...,m$).\\ 
Proof: {\rm Let $z_1,\ldots ,z_{2m}$ be the eigenvalues of $T(E)$. The
coefficients 
$$ a_k=(-1)^k \sum_{i_1<\ldots <i_k} z_{i_1}\cdots z_{i_k},\qquad  k=1\ldots m, $$
can be expressed as combination of traces of powers of $T(E)$ of degree $k$:
$a_1=-\tr\,T(E)$, $a_2=\frac{1}{2}[\tr\, T(E)]^2-\frac{1}{2}\tr\,[T(E)^2]$, 
etc. Since $T(E)=E^nT_n+\ldots +T_0$, the coefficient
$a_k$ is a polynomial of degree $kn$ in $E$. The remaining 
coefficients $a_{2m-k}$ are discussed differently. The point is that 
$a_{2m}=z_1\cdots z_{2m}=\det T(E)$ is independent of $E$ and the 
coefficients can be written as 
$$ a_{2m-k}=(-1)^k \sum_{i_1<\ldots <i_{2m-k}} z_{i_1}\cdots z_{i_{2m-k}} =
(-1)^k a_{2m}\sum_{i_1<\ldots <i_k} (z_{i_1}\cdots z_{i_k})^{-1} $$
Therefore, $a_{2m-1}=-a_{2m}\,\tr [T(E)^{-1}]$,
$a_{2m-2}= a_{2m}\frac{1}{2} [\tr  T(E)^{-1}]^2 -a_{2m}\frac{1}{2} 
\tr [T(E)^{-2}]$, etc. Since also $T(E)^{-1}$ 
is a polynomial matrix of degree $n$ in $E$, $a_{2m-k}$ is a polynomial
of degree $kn$ in $E$.}\hfill $\square $
\end{prop}

\begin{thrm}[{\bf Duality}]{\qquad}
\begin{equation}
\det [z\mathbb I_{2m}-T(E)]= (-z)^m  
\frac{\det [E\mathbb I_{nm}- H(z)]}{\det(B_1\cdots B_n)}    
\end{equation}
\noindent
{\it Proof}: {\rm According to proposition~\ref{powerexpT} 
the leading term in the expansion in $E$ of 
$\det \left[ z\mathbb I_{2m} - T(E)\right ]$ coincides with the leading 
term in the expansion of $\det [z\mathbb I_{2m}-E^nT_n]$, which is 
$(-z)^mE^{nm} \det [B_1\cdots B_n]^{-1}$. The 
leading term of  $\det [E\mathbb I_{2m}-H(z)]$ is $E^{nm}$. Since 
by proposition~\ref{statduality} the two polynomials, for given $z$, have the 
same zeros in $E$, they must be proportional by a constant.} \hfill $\square $
\end{thrm}

\begin{figure}\label{spctrl}
\begin{center}
\includegraphics[angle=0,width=6cm]{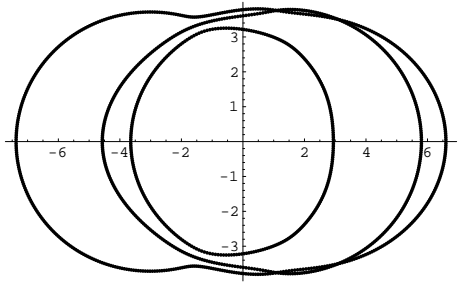}
\caption{\label{Fig2} 
The eigenvalues of the Hamiltonian matrix ($n=8, m=3$) of a 2D Anderson 
model on a lattice $3\times 8$, diagonal disorder parameter $w=7$. 
The b.c. parameter is $z=\exp (n \xi + i\varphi)$ with $\xi= 1.5$.
As $\varphi $ changes, the 24 eigenvalues (of the balanced matrix) trace 
$m=3$ closed loops of equation $\ln |z_k(E)|=1.5 $.}
\end{center}
\end{figure}

This relation among characteristic polynomials
is a ``duality identity'' as it exchanges the roles of the parameters 
$z$ and $E$ among the two matrices: $z$ is an eigenvalue of $T(E)$ 
if and only if $E$ is 
an eigenvalue of the block tridiagonal matrix $H(z)$. I gave different 
proofs of it \cite{Molinari97,Molinari98,Molinari08}. With $z=1$ it is
a tool for computing determinants of block tridiagonal or banded matrices
with corners.\\
The eigenvalues of $H(z)$ make the l.h.s. of duality equal to zero, i.e. 
there is at least a complex factor $z_i(E)-z=0$. This means that an 
eigenvalue $E$ is at the intersection of a line $|z_i(E)|=|z|$ and 
arg $z_i(E)=\,$arg $z$. By changing only the parameter arg $z$, the 
eigenvalues move along {\em spectral lines} $|z_i(E)|=|z|$.
For tridiagonal matrices ($m=1$) there is a single spectral curve 
(figure~\ref{Fig1}), for $m>1$ several 
spectral curves appear \cite{Molinari09} (see figure~\ref{Fig2}).   

A more symmetric duality relation results from multiplication of
the dual identities for $(T-z)$ and $(T-1/z)$:
\begin{equation*}
\fl 
\det\left [T(E)+T(E)^{-1} - \left( z+\frac{1}{z}\right)\mathbb I_{2m}\right] = 
\frac{\det [E \mathbb I_{nm}-H(z)] \det[ E \mathbb I_{nm} - H(1/z)]}
{\det [B_1\cdots B_n]\det[C_1\cdots C_n]}
\end{equation*}

\section{Transfer matrix and resolvent}

\Eref{chain} with open b.c. $u_0=0$ and $u_{n+1}=0$, is the eigenvalue
equation for the matrix 
\begin{equation}
h =\left[\begin{array}{cccc}
A_1 & B_1 & {} &  {} \\
C_2  & \ddots & \ddots & {}\\
{} & \ddots & \ddots & B_{n-1}\\
{} & {} & C_n & A_n 
\end{array}\right]
\end{equation}
Let  $(u_1,\ldots , u_n)^t$ be a (right) eigenvector of $h$ with 
eigenvalue $E$; then $u_1$ and $u_n$ are both nonzero, or the 
whole vector would be null by the chain recursion. With the block partition
\begin{equation*}
 T(E) = \left[ \begin{array}{cc} T(E)_{1,1} & T(E)_{1,2}\\ 
T(E)_{2,1} & T(E)_{2,2}
\end{array}\right ] 
\end{equation*}
\eref{transfer} gives $T(E)_{1,1}u_1=0$ and 
$T(E)_{2,1}u_1=u_n$. This means that $\det T(E)_{1,1}=0$ whenever 
$\det [E\mathbb I_{nm} - h]=0$ (and $\det T(E)_{2,1}\neq 0$).
The following duality relation results:
\begin{prop}[{duality for the open chain}]{\qquad}
\begin{equation}
 \det [E\mathbb I_{nm} - h]\;= \,\det T(E)_{1,1} \,\det[B_1\cdots B_n] 
\end{equation}
Proof: {\rm
by construction $T(E)_{11}=E^n(B_1\cdots B_n)^{-1}+$ lower powers in $E$. Then
both $\det [E\mathbb I_{nm}-h]$ and $\det T(E)_{11}$ are polynomials in $E$ 
of degree $nm$.
Having the same roots, they are proportional.} \hfill $\square $
\end{prop} 

The blocks $T(E)_{12}$ and $T(E)_{21}$ are polynomial matrices of
degree $n-1$ in $E$, and $T(E)_{22}$ has degree $n-2$.
The four blocks can be evaluated in terms 
of the corner blocks of the resolvent matrix 
\begin{equation*}
  g(E)=[h-E\mathbb I_{nm}]^{-1} =: \left[\begin{array}{ccc}
g_{1,1} & \cdots & g_{1,n}\\
\vdots & {} & \vdots \\ 
g_{n,1} & \cdots & g_{n,n}\end{array}\right]  
\end{equation*}
The corner matrices $C_1$ and $B_n$ are absent in $h$ but enter in the 
definition of $T(E)$ through the 1-step factors $t_1(E)$ and $t_n(E)$, 
and will be accounted for.
\begin{prop}
Let $g_{i,j}\in \mathbb C^{m\times m}$ $(a,b=1\ldots n)$ be the blocks of 
$g(E)$. Then
\begin{eqnarray}\label{Tg}
T(E)= \left[\begin{array}{cc} 
-B_n^{-1}(g_{1,n})^{-1} & -B_n^{-1}(g_{1,n})^{-1}g_{1,1}C_1\\ 
g_{n,n}(g_{1,n})^{-1} &g_{n,n}(g_{1,n})^{-1}g_{1,1}C_1  -g_{n,1}C_1
\end{array}\right]
\end{eqnarray}
Proof: {\rm write the identity 
$[h-E\mathbb I_{nm}]g(E) =\mathbb I_{nm}$ for the block indices 
$i=2\ldots n-1$ and $k=1,n$:
$ C_ig_{i-1,k}+(A_i -E\mathbb I_m) g_{ik}+B_i g_{i+1,k}= 0 $.
The recursive relations are solved by the transfer matrix method
and give a matrix relation among the corner blocks:
\begin{eqnarray*}
\left[\begin{array}{cc} g_{n,1}& g_{n,n}\\ g_{n-1,1} & g_{n-1,n} \end{array}
\right]
=t_{n-1}(E)\cdots t_2 (E)\left[\begin{array}{cc}g_{2,1}& g_{2,n}\\g_{1,1} 
& g_{1,n}\end{array}\right]
\end{eqnarray*}
Left multiply both sides by $t_n(E)$ and simplify l.h.s. by means of the 
identity
$C_ng_{n-1,k}+(A_n-E\mathbb I_m)g_{n,k}=\delta_{k,n}\mathbb I_m$. Insert 
$t_1(E)t_1(E)^{-1}=\mathbb I_{2m}$ in the r.h.s. to obtain $T(E)t_1^{-1}$, 
and simplify the 
action of $t_1^{-1}$ by means of the identity $(A_1-E\mathbb I_m)g_{1,k}+
B_1 g_{2,k}=\delta_{1,k}\mathbb I_m$. The useful factorization is obtained:
\begin{eqnarray}\label{m=Tm}
\left[\begin{array}{cc} 0 & -B_n^{-1}\\ g_{n,1} & g_{n,n} \end{array}\right]
=T(E)\left[\begin{array}{cc}g_{1,1}& g_{1,n}\\-C_1^{-1} & 0\end{array}\right]
\end{eqnarray}
A matrix inversion and multiplication give the result}.\hfill $\square $
\end{prop}
The representation provides the transfer matrix through a large matrix 
inversion, rather than multiplications. It was employed by Kramer and 
MacKinnon \cite{Kramer} in a numerical proof of one-parameter 
scaling for the localization length of Anderson's model.

\section{Exponential inequalities} 
Products of random matrices are known to exhibit Lyapunov exponents
that are asymptotically stable and self-averaging, i.e. independent of the 
length $n$ and of the realization of the random product.
In the present deterministic approach a single chain is 
considered, and it will be shown that it is possible to 
give exponential bounds on the eigenvalues for long chains, that 
justify the introduction of exponents.

Demko, Moss and Smith \cite{Demko} made the very general statement that,
loosely speaking, the matrix elements of the inverse of block tridiagonal or
banded matrices {\em decay exponentially from the diagonal} (see 
also \cite{Benzi, Meurant}). I here present their interesting proof 
adapted to the block partitioning of matrices. I then apply it to the 
matrix $g(E)$ to obtain bounds for the singular values of $T(E)$.\\

The main ingredient is the best approximation of the function $(x-a)^{-1}$ 
on the interval $[-1,1]$ ($|a|>1$) by a polynomial of degree  
$k$, which was obtained by Chebyshev together with the determination 
of the error \cite{Meinardus}. With proper rescaling it is \cite{Demko}:
\begin{lem}
Let $P_k$ be the set of real monic polynomials of degree $k$, let
$[a,b]$ be an interval of the positive real line, with $a>0$. Then:
\begin{eqnarray}
\inf_{p_k\in P_k} \left \{ \sup_{x\in [a,b]}\left | \frac{1}{x}-p_k(x)\right |
\right \} = \,C\; q^{k+1},\\ 
C= \frac{(\sqrt b +\sqrt a)^2}{2ab}, 
\qquad q= \frac{\sqrt b -\sqrt a}{\sqrt b+ \sqrt a} \label{Cq}
\end{eqnarray}
\end{lem}
If $A$ is a block tridiagonal matrix with blocks of size $m\times m$ and
if $p_k(x)$ is a polynomial of degree $k$, the blocks $p_k(A)_{i,j}$ 
of the matrix $p_k(A)$ are null for $|i-j|>k$. \\
Let $A$ be a positive definite block tridiagonal matrix, with inverse $A^{-1}$.
If $A^{-1}[i,j]$ denotes any matrix element in the block $(A^{-1})_{ij}$ 
then, for any monic real polynomial of degree $k= |i-j|-1$, it is:
\begin{eqnarray*}
|A^{-1}[i,j]| &=   \left | A^{-1}[i,j]-p_k(A)[i,j]\right | \\
&\le  \| A^{-1}-p_k(A)\|
=  \sup_{\lambda\in sp(A)} \left | \frac{1}{\lambda} - p_k(\lambda )\right |\\
&\le  \sup_{\lambda\in [a,b]} \left | \frac{1}{\lambda} - p_k(\lambda )\right | 
\end{eqnarray*}
where $\| A \|=\sup_{\|x\|=1} \|Ax\|$ is the operator 
norm\footnote{For any matrix $A$ with matrix
elements $A_{rs}$ it is $|A_{rs}|=|(e_r|Ae_s)|\le \|Ae_s\|\le \|A\| $, 
where $e_i$ are canonical unit vectors and Schwarz's inequality is used.}, 
and the spectral theorem is used. 
In the last line $[a,b]$ is the smallest interval containing the spectrum
of eigenvalues $sp(A)$.
Next, the $\inf $ is taken over the polynomials $p_k$. The lemma states that
the minimum exists, and the error gives the main inequality.
Note that for $|i-j|=0$: $|A^{-1}[i,i]|\le \|A^{-1}\|= 1/a$. Therefore:
\begin{thrm}[Demko, Moss and Smith]\label{Demko1}
Let $A$ be a positive definite block tridiagonal matrix, with square blocks 
of size $m$, let $[a,b]$ be the smallest interval containing the spectrum of 
$A$, let $A^{-1}[i,j] $ be any matrix element in the block 
$(A^{-1})_{ij}$. Then:
\begin{equation}
\left |A^{-1}[i,j] \right | \le \cases{C\, q^{|i-j|} & for  $|i-j|\ge 1$\\
1/a & for $i=j$\\}
\end{equation}
where $q < 1$ and $C$ are specified by eq.\eref{Cq}.
\end{thrm}

Demko et al. also proved an extension of the theorem to a matrix $A$ 
that is block tridiagonal invertible but fails to be positive. 
An estimate for $A^{-1}$ is obtained by noting that
$A^{-1}= A^\dagger (AA^\dagger)^{-1}$.
The matrix $AA^\dagger $ is block 5-diagonal positive definite, and
a polynomial $p_k(AA^\dagger)$ is a matrix whose blocks $(i,j)$ are null if
$|i-j|>2k$. The previous theorem applies, 
with $[a,b]$ being the smallest positive interval containing 
$sp(AA^\dagger)$:
$$ |(AA^\dagger)^{-1}[i,j]| \le \frac{C}{\sqrt q} 
\, q^{\frac{1}{2}|i-j|},\quad |i-j|>2 $$
The extension of the theorem is here written in the block notation, with
minor changes from the original paper:
\begin{thrm}\label{Demko2}
Let $A$ be an invertible block tridiagonal matrix with square blocks 
of size $m$, let $[a,b]$ be the smallest interval containing 
$sp (A^\dagger A)$, let $A^{-1}[i,j] $ be any matrix element in the block 
$(A^{-1})_{ij}$. Then:
\begin{eqnarray}
|A^{-1}[i,j] | \le C_i\, q^{\frac{1}{2}|i-j|}\\
C_i = \frac{C}{q}  \left ( \|A_{i-1,i}\| +\|A_{i, i}\| +\|A_{i+1,1}\| 
\right ),
\end{eqnarray}
where $q<1$ and $C$ are given in \eref{Cq}.\\
Proof: {\rm in terms of block multiplication:
\begin{equation*} 
\fl (A^{-1})_{ij} =  (A^\dagger)_{i,i-1}[(AA^\dagger)^{-1}]_{i-1,j}+
(A^\dagger)_{i,i} [(AA^\dagger)^{-1}]_{i,j}+ (A^\dagger)_{i,i+1}
[(AA^\dagger)^{-1}]_{i+1,j}.
\end{equation*}
The sup norm, the triangle inequality, the property
$\|AB\|\le \|A\| \|B\|$, and the bound on $(AA^\dagger)^{-1}$ give:
\begin{eqnarray*}
\fl \|(A^{-1})_{ij} \| \le  \| A_{i-1,i} \| \|(AA^\dagger)^{-1}{}_{i-1,j} \|
 + \| A_{i,i} \| \|(AA^\dagger)^{-1}{}_{i,j}\| 
+ \| A_{i+1,i} \| \|(AA^\dagger)^{-1}{}_{i+1,j} \| \\
\le  \frac{C}{\sqrt q}\left( \| A_{i-1,i} \|  q^{\frac{1}{2}|i-j-1|}
 + \| A_{i,i} \|  q^{\frac{1}{2}|i-j|}
+  \| A_{i+1,i} \|  q^{\frac{1}{2}|i-j+1|}\right ) \\
\le \frac{C}{q}  \left ( \|A_{i-1,i}\|+\|A_{i,i}\|+\|A_{i+1,j}\|\right ) 
\; q^{\frac{1}{2}|i-j|} 
\end{eqnarray*}
If $A^{-1}[i,j]$ is any matrix element in 
the block $(A^{-1})_{ij}$, it is 
$\left | A^{-1}[i,j] \right | \le  \|A_{ij}\| $.}
\hfill $\square $
\end{thrm}

Given an invertible matrix $A$, the {\em condition number} of $A$ is
\cite{Bhatia}:
\begin{equation*}
\rm{cond}\; (A)=:\|A\|\, \|A^{-1}\| 
\end{equation*}
In general it is cond $(A) \ge 1$. If $a$ and $b$ are the extrema 
of the spectrum of a positive matrix $P$ it is $b=\|P\|$ and 
$1/a=\|P^{-1}\|$; then $b/a =$ cond $(P)$.\\
Since $\|AA^\dagger \|=\|A\|^2$, it is cond $(AA^\dagger )=
[\rm{cond}(A)]^2$ and the parameters in theorem \ref{Demko2} are:
\begin{equation}
 q=\frac{\rm{cond}(A) -1}{\rm{cond}(A)+1} , \qquad
C=\frac{(\rm{cond}(A)+1)^2}{2\|A\|^2} 
\end{equation}
{\quad}\\

Theorem \ref{Demko2} is applied to the corner blocks of the 
resolvent $g(E)=[h-E\mathbb I_{nm}]^{-1}$, $E\notin sp(h)$,
which enter in the representation \eref{Tg} of the transfer matrix. 
The number cond $(h-E)$ defines the parameters $q<1$ and $C$.
\begin{prop}
If $g[1,n]$ and $g[n,1]$ are matrix elements of the corner blocks 
$g_{1n}$ and $g_{n1}$ of $g(E)$, then the following inequalities hold:
\begin{eqnarray}
|g[1,n] | \;\le \; C\, (\|A_1-E\|+ \|B_1\|)  q^{\frac{1}{2}(n-3)},\\
|g[n,1] | \;\le \; C\, (\|A_n-E\|+ \|C_n\|)  q^{\frac{1}{2}(n-3)}
\end{eqnarray}
where $A_1,B_1,A_n,B_n$ are the blocks in the first and last row of $h$.
\end{prop}
We prepare for the main theorem with the following lemma:
\begin{lem}
The singular values $\theta_k$ of the block $T_{11}$ of $T(E)$ are 
exponentially large in $n$: $\theta_k >  q^{-n/2}/K$.\\
Proof: {\rm from $(T_{11})^{-1}= -g_{1n}B_n$ it follows that: 
$ \tr [(T_{11}^\dagger T_{11})^{-1}]=
\tr [B_nB_n^\dagger g_{1n}^\dagger g_{1n}]\le m^2\|B_nB_n^\dagger\|
\|g_{1n}^\dagger g_{1n}\|=m^2\|B_n\|^2\|g_{1n}\|^2\le
m^2\|B_n\|^2 C^2 (\|A_1-E)\|+ \|B_1\|)^2 q^{n-3}=: K^2\, q^n $.
Since $ \tr [(T_{11}^\dagger T_{11})^{-1}]=\sum_{k=1}^m \theta_k^{-2}$, 
it turns out that each singular value of $T_{11}$ is 
larger than $q^{-{n/2}}/K$}.\hfill $\square $
\end{lem}


\begin{mainthrm}
If $q<1$ and $n$ is large, the transfer matrix $T(E)$ has $m$ singular values
larger than $\frac{1}{K}q^{-n/2}$ and $m$ singular values smaller than 
$K q^{n/2}$.\\
Proof: {\rm
Let $\theta_1\ge \ldots \ge \theta_m$ be the singular values of the block 
$T_{11}$, and let $\sigma_1\ge \ldots \ge\sigma_{2m}$ be the singular values 
of $T(E)$. The interlacing property (Theorem 7.12 of ref.\cite{ZhangF}) 
states that:
$$\sigma_{k}\ge \theta_k \ge \sigma_{m+k}, \quad k=1,\ldots ,m $$ 
Therefore, there are at least $m$ singular values of $T(E)$ that are larger 
than $\frac{1}{K}q^{-n/2}$. Since the same conclusion holds true for 
$T(E)^{-1}$, which is 
similar to a transfer matrix by proposition \ref{prop1}, 
there are precisely $m$ 
singular values of $T(E)$ that are larger than $\frac{1}{K}q^{-n/2}$, 
and $m$ that are smaller than $K q^{n/2}$}.\hfill $\square $
\end{mainthrm}

\rem{++++++++++++++++++++
The statement that the polynomial $\det T(E)_{11}$ 
is exponentially large in $n$ for $E$ away from a zero of $\det [h-E]$, 
is a fact explained with rare beauty and generality by Cartan's Lemma:
\begin{lem} [Cartan]
Let $z_1\ldots z_n$ be given points in $\mathbb C$ and $H>0$ be given.
Then there exist closed disks $\Delta_1\ldots \Delta_m$ ($m\le n$) such that
the sum of their radii is less that $2H$ and
\begin{equation}
|z-z_1| |z-z_2|\cdots |z-z_n| > \left( \frac{H}{e}\right )^n
\end{equation}
whenever $z$ does not belong any of the disks.
\end{lem}
++++++++++++++++}

\section{Jensen's formula and the exponents} 
The two sides of the duality relation are determinantal expressions of the same 
polynomial in two variables, $F(z ,E) =: \det \,[z \mathbb I_{2m}-T(E)]$.
Let $z_1, \ldots ,z_{2m}$ be the zeros in the variable $z$ (the 
eigenvalues of $T(E)$) with $|z_1|\ge \ldots \ge |z_{2m}|$, and 
let $E_1,\ldots ,E_{nm}$ be the zeros in $E$ (the eigenvalues of $H(z)$). 
It is convenient to introduce
the {\em exponents} of the transfer matrix:
$$   \xi_k =: \frac{1}{n}\ln |z_k|  $$

\begin{remark}\label{expvslyap}
The exponents are not to be confused with the Lyapunov exponents, which are
defined in terms of the positive eigenvalues $x_k$ of the matrix $T^\dagger T$. 
It has been shown (with less general $T$) that also $T^\dagger T$ is the 
transfer matrix of a block tridiagonal matrix, so the same discussion may 
be applied to them \cite{Molinari09}.
\end{remark}
 
The sum of the exponents is $\frac{1}{n}\ln |\det T(E)|$. Then:
\begin{equation}
\sum_{k=1}^{2m}\xi_k = \frac{1}{n}\sum_{j=1}^n \left(
\ln |\det C_j|-\ln |\det B_j|\right )\label{sumallxi}
\end{equation}
Some general analytic results are now given, based on the following theorem
of complex anaysis \cite{Markushevic}:
%
\begin{thrm}[Jensen]
If $f$ is holomorphic and $f(0)\neq 0$, and $z_1\ldots z_n$ are its 
zeros in the disk of radius $r$, then:
$\int_0^{2\pi} \frac{d\theta}{2\pi} \ln |f(re^{\rmi\theta})| =  \ln |f(0)|-
\sum_k \ln (|z_k|/r).$ 
\end{thrm}

\noindent
The theorem is applied to $F(z,E)$ as a function of $z$, resulting in a 
relation between a sum of the exponents and the spectrum of the Hamiltonian 
matrix \cite{Molinari03}:

\begin{prop}
\begin{eqnarray}
\fl \frac{1}{m} 
\sum_{\xi_k <\xi} (\xi-\xi_k) \; -\xi \nonumber\\
 =\frac{1}{mn}\int_0^{2\pi}\frac{d\varphi}{2\pi}\ln \left |
\det [H(\exp [n\xi+\rmi\varphi ])-E] \right |
-\frac{1}{mn}\sum_{j=1}^n\ln \left | \det\,C_j \right |\label{dual2}
\end{eqnarray}
\noindent
Proof: {\rm 
Jensen's theorem with $z=e^{n\xi +\rmi\theta}$ gives in the r.h.s. the 
sum of exponents contained the disk of radius $e^{n\xi}$:
$$\int_0^{2\pi}\frac{d\theta}{2\pi} \ln \left |
F(e^{n\xi+\rmi\theta},E)\right | = 
\ln |\det T(E)| + n\sum_{k=1}^{2m} (\xi-\xi_k) \theta (\xi-\xi_k).$$
The dual expression is used in the l.h.s.: $\ln |F|= mn\xi +
\ln |\det [H(e^{n\xi+\rmi\varphi})-E]|-\sum_j\ln |\det B_j|.$} \hfill $\square $ 
\end{prop}

A derivative in the variable $\xi $ of \eref{dual2} gives the counting 
functions of exponents  $N(\xi,E) = \sum \theta (\xi-\xi_a(E))$, which is
also obtainable by Euler's formula for the zeros $z_k$ of the $F(z,E)$ 
\cite{Molinari11}. 

Hadamard-Fisher's inequality \cite{Bhatia,ZhangF} 
states that if $M_1, \ldots , M_n$ are the diagonal blocks of the 
positive matrix $A^\dagger A$, then $|\det A|^2 \le \det M_1 \cdots \det M_n$.\\
The inequality is applied to the r.h.s. in eq.\eref{dual2}, with the 
balanced matrix $H^B(e^{\xi+\rmi\varphi/n})$:
\begin{eqnarray}
\fl \sum_{k=1}^{2m} ( \xi -\xi_k) \theta(\xi-\xi_k) - m\xi 
\le -\frac{1}{n}\sum_{j=1}^n\ln \left | \det\,C_j \right | \\
 +\frac{1}{2n}\sum_{k=1}^n \ln \det \left [ (A_k^\dagger -\overline E)
(A_k-E) + e^{2\xi} B_k^\dagger B_k + e^{-2\xi} C_k^\dagger C_k \right ]
\nonumber
\end{eqnarray}
If the norms of matrices $A_i$ $B_i$ and $C_i$ are bounded by some constant 
for all $i$, and $m$ is fixed, 
the sum in l.h.s. of inequality remains finite for any length $n$, as 
the r.h.s. is an average value for the blocks. 
 
\begin{cor} The sum of the positive exponents is obtained from \eref{dual2} 
with $\xi=0$ and by means of eq.\eref{sumallxi}
\begin{eqnarray}
\fl \sum_{k=1}^{2m} \xi_k\, \theta(\xi_k)
=\frac{1}{n}\int_0^{2\pi}\frac{d\theta}{2\pi}\ln \left |
\det [H(e^{\rmi\theta})-E]\right | -\frac{1}{n}\ln |\det\,[B_1\cdots B_n ] |
\label{thoulesslike}
\end{eqnarray}
\end{cor}
The identity is exact and applies to a single transfer
matrix. It is reminiscent of the formula \eref{KuSoTh} 
for the sum of the Lyapunov exponents of random transfer matrices.
The ``angular average'' replaces the ensemble averaged density of 
eigenvalues $\rho(E)$, which was extended to tridiagonal non-Hermitian 
matrices in \cite{Derrida, Goldsheid05}.

\section{The Hermitian difference equation}
Most of the literature concentrates on the Hermitian case. However,
as duality requires $z$ to be a complex parameter, the matrix $H(z)$ fails 
to be Hermitian unless $|z|=1$;
\begin{equation}
H (z)=\left[\begin{array}{cccc}
A_1 & B_1 & {} &  \frac{1}{\displaystyle z}B_n^\dagger \\
B_1^\dagger  & \ddots & \ddots & {}\\
{} & \ddots & \ddots & B_{n-1}\\
z B_n & {} & B_{n-1}^\dagger & A_n 
\end{array}\right], \qquad A_k=A_k^\dagger
\end{equation}
A useful symplectic property holds for
the transfer matrix (in transport problems it describes flux conservation,
\cite{Molinari97}), and implies that exponents come in pairs $\pm \xi_a $:
\begin{prop}
\begin{equation}
T(\overline E)^\dagger \Sigma_n T(E)=\Sigma_n, \qquad
\Sigma_n = 
\rmi\left[\begin{array}{cc} 0 & -B_n^\dagger \\ B_n & 0\end{array}\right]
\nonumber
\end{equation}
\noindent
Proof: {\rm in the factorization 
$T(E)=t_n(E)\cdots t_1(E)$, the factors $t_k(E)$ 
($k=2\ldots n$) have the property 
$t_k(\overline E )^\dagger \Sigma_k t_k(E)=\Sigma_{k-1}$. The factor $t_1$
that contains the boundary blocks, closes the loop:  
$t_1(\overline E )^\dagger \Sigma_1 t_1(E)=\Sigma_n$}.\hfill $\square $
\end{prop}
%
%
\begin{cor} If $E$ is real, the eigenvalues of $T(E)$ different from $\pm 1$
come in pairs $z$, $1/\overline z$. The associated exponents are opposite.\\
Proof: {\rm If $T(E)u=zu$, the symplectic property implies that 
$T(E)^\dagger\Sigma_nu=1/z \Sigma_nu$ i.e. $1/\overline z$ is an eigenvalue
of $T(E)$. Moreover, if $|z|\neq 1$, then $u^\dagger \Sigma_n u=0$.}
\end{cor}

\begin{prop}
If Im $E\neq 0$ then $T(E)$ has no eigenvalues on the unit circle.\\
Proof: {\rm for Im $E\neq 0$ and $z=e^{\rmi\theta}$ it is always 
$\det [E-H(e^{\rmi\theta})]\neq 0$ because $H(e^{\rmi\theta})$ is Hermitian and 
it has real eigenvalues. Therefore, by duality, 
$\det [T(E)-e^{\rmi\theta}\mathbb I_{2m}]$ never vanishes.}\hfill $\square $
\end{prop}

A degeneracy occurs in the exponents of the real transfer matrix of a real 
symmetric difference equation (the Anderson model is a notable example,
but remind remark \ref{expvslyap}):

\begin{prop}
Let the matrices $A_k$ be real symmetric and $B_k$ be real invertible.
For $E\in \mathbb R$, the real eigenvalues of $T(E)$ come in pairs $z$, $1/z$,
the complex ones also have the
conjugated pair $\overline z$, $1/\overline z$.\\
Proof: {\rm if $z$ is a complex eigenvalue of $T(E)$ not in the unit circle,
then also $\overline z$, $1/\overline z$ and $1/z$ are distinct 
eigenvalues, and exponents are doubly degenerate opposite pairs.
If $z$ is a real eigenvalue, then $1/z$ is an eigenvalue. Therefore 
an eigenvalue (real or not) is always paired to the eigenvalue $1/z$.}
\hfill $\square $
\end{prop}

\section*{Acknowledgement}
I wish to dedicate this work to prof. Giovanni Cicuta, in his 70th birthday,
as a sign of gratitude.

\vfill
\end{document}